\documentclass[preprint,prb]{revtex4}
\usepackage{amsmath}
\usepackage{amssymb}
\usepackage{amsthm}

\def\v#1{{\bf#1}}

\begin{document}

\title{Comment on ``Helmholtz theorem and the v-gauge in the problem of superluminal and instantaneous signals in classical electrodynamics," by A. Chubykalo et al}
\author{Jos\'e A. Heras}
\email{heras@phys.lsu.edu}
\affiliation{Departamento de F\'\i sica, E. S. F. M., Instituto
Polit\'ecnico Nacional, M\'exico D. F. M\'exico and Department of Physics and Astronomy, Louisiana State University, Baton
Rouge, Louisiana 70803-4001, USA}
\begin{abstract}
Fundamental errors in the Chubykalo et al paper [2] are highlighted. Contrary to their claim that ``... the irrotational component of the electric field has a physical meaning and can propagate exclusively {\it instantaneously}," it is shown that this instantaneous component is physically irrelevant because it is always canceled by a term contained into the solenoidal component. This result follows directly from the solution of the wave equation 
that satisfies the solenoidal component. Therefore the subsequent inference of these authors that there are two mechanisms of transmission of energy and momentum in classical electrodynamics, one retarded and the other one instantaneous, has no basis. The example given by these authors in which the full electric field of an oscillating charge equals its instantaneous irrotational component on the axis of oscillations is proved to be false. 
\end{abstract}

\maketitle

\noindent{\normalsize {\bf 1. INTRODUCTION}}

\bigskip
\noindent
In the context of the so-called velocity gauge (v-gauge) of classical electrodynamics [1], in which the scalar potential $\varphi$ and the vector potential $\v A$ satisfy the condition: $\nabla\cdot \v A+[c/v^2]\partial\varphi/\partial t=0,$
and applying the Helmholtz theorem, which states that $\v E=\v E_i+\v E_s$ where $\v E_i$ is irrotational $(\nabla\times\v E_i=0)$ and $\v E_s$ is solenoidal $(\nabla\cdot\v E_s=0)$, Chubykalo et al [2] make the claim that ``...the irrotational component of the electric field has a physical meaning and can propagate exclusively {\it instantaneously}." These authors also claim that ``... there are {\it two} mechanisms of the energy and momentum transmission in classical electrodynamics: (1) the {\it retarded} one by means of a radiation $(\v E_s$ and $\v B)$...; (2) the {\it instantaneous} one by means of the irrotational field $\v E_i$." To illustrate their conclusion (2), they discuss the problem of an oscillating charge and claim that in this case the full electric field is equal to its irrotational component on the axis of oscillations. In other words,  they claim to have found a full electric field  $\v E$ of Maxwell's theory satisfying the equation $\v E =\v E_i$ where $\v E_i$ is an instantaneous field! 

The purpose of this comment is to point out that the above claims are incorrect. This means that the instantaneous component $\v E_i$ has no physical meaning and therefore there is no two transmission mechanisms in the electrodynamics of localized sources in vacuum. We point out that the misleading conclusions of Chubykalo et al [2] arise from having done a separated lecture of the equations for the components $\v E_i$ and $\v E_s$, {\it i.e.,} these authors treat these equations as if they were independent equations. But the fact is that they are coupled and therefore no physical inference on $\v E_i$ and $\v E_s$ should be extracted from these equations without considering their coupling. 

Here we show that the component $\v E_i$ is physically irrelevant because it is always canceled by its additive negative $-\v E_i$ which is contained into the solenoidal component $\v E_s$. The fact that $-\v E_i$ is a part of 
$\v E_s$ follows directly from the solution of the wave equation that satisfies $\v E_s$. Chubykalo et al [2] have also ignored the comment of Jackson [3] on a paper of Chubykalo and Vlaev [4], in which Jackson has emphasized that the inference of both electromagnetic interactions, instantaneous and retarded ones, cannot be made. 

The example of Chubykalo et al [2] in which for an oscillating charge they conclude that $\v E=\v E_i$ on the axis of oscillations is proved to be false in the Maxwell theory, but as a consolation for these authors this result is shown to be correct in the context of a Galilei-invariant electromagnetic theory [5,6]. Furthermore, not only does Chubykalo et al [2] paper display fundamental errors but it misinforms the readers by attributing to me an incorrect inference on the decomposition of a retarded field that I have never made.
\bigskip
\bigskip

\noindent{\normalsize {\bf 2. TWO KINDS OF ELECTRIC FIELD?}}

\bigskip
\noindent Starting with the v-gauge potentials and applying the Helmholtz theorem to: (i) the v-gauge vector potential 
$\v A=\v A_i+\v A_s$; (ii) the retarded electric field $\v E=\v E_i+\v E_s$ and (iii) the current density $\v j=\v j_i+\v j_s$, Chubykalo et al [2] derive the following equations 
\begin{eqnarray}
\nabla^2\v E_i=4\pi\nabla\rho,\;\\
\nabla^2\v E_s-\frac{1}{c^2}\frac{\partial^2\v E_s}{\partial t^2}=\frac{4\pi}{c^2}\frac{\partial\v j_s}{\partial t},
\end{eqnarray}
[see Eq. (24) and the line below Eq. (23) both in Ref. 2]. 
They claim: ``Thus we see that the vector fields $\v E_i$ and $\v E_s$ are solutions of {\it different} equations with $\v E_i$-``wave" propagating {\it instantaneously} and $\v E_s$-wave propagating with the velocity $c$ respectively." 
Clearly, these conclusions are obtained from considering separately Eqs. (1) and (2).

Furthermore, without making use of the v-gauge potentials, the authors of Ref. 2 also apply the Helmholtz theorem directly to both the wave equation of the electric field and the Ampere-Maxwell equation and obtain again Eqs. (1) and (2) [Eqs. (27) and (28) of Ref. 2]. In fact, it is not difficult to show that the equations 
\begin{eqnarray}
\nabla^2\v (\v E_i+\v E_s) -\frac{1}{c^2}\frac{\partial^2(\v E_i+\v E_s)}{\partial t^2}=4\pi\Bigg(\nabla\rho+\frac{1}{c^2}\frac{\partial\v (\v j_i+\v j_s)}{\partial t}\Bigg),\\
\nabla\times(\v B_i+\v B_s) -\frac{1}{c}\frac{\partial(\v E_i+\v E_s)}{\partial t}=\frac{4\pi}{c}(\v j_i+\v j_s),\qquad\qquad\qquad
\end{eqnarray}
imply Eqs. (1) and (2) as well as the equation
\begin{eqnarray}
-\frac{1}{4\pi}\frac{\partial\v E_i}{\partial t}= \v j_i,
\end{eqnarray}
[Eq. (30 of Ref. 2]. 
First of all, we note that Eqs. (1) and (2) are {\it different} but not {\it independent.} Actually, they are coupled equations. From a formal point of view Eq. (1) states that the field $\v E_i$ propagates instantaneously, but before concluding that this acausal feature of $\v E_i$ is a physical prediction of Maxwell's theory, we should consider also Eqs. (2) and (5) because the latter involves explicitly $\v E_i$ and the former involves implicitly 
$\v E_i$ via the current $\v j_s=\v j-\v j_i$.  In fact, using this decomposition of the current together with Eq. (5) we can write Eq. (2) as
\begin{eqnarray}
\nabla^2\v E_s-\frac{1}{c^2}\frac{\partial^2\v E_s}{\partial t^2}=\frac{4\pi}{c^2}\frac{\partial\v j}{\partial t}+\frac{1}{c^2}\frac{\partial^2\v E_i}{\partial t^2}.
\end{eqnarray}
It is now clear that (1) and (6) are coupled equations. From Eq. (6) we see that the component $\v E_i$ may be considered as a source of the component $\v E_s$ and therefore it would not be surprising that the solution of Eq. (6) would involve information on $\v E_i$. This means that the statement that 
the solenoidal component $\v E_s$ propagates with speed $c$ is simplistic because one of the sources in Eq. (6), namely, $\v E_i$ extends over all space and propagates instantaneously. 
In order to find what is the exact connection between $\v E_s$ and $\v E_i$ predicted by Eq. (6) [or equivalently by Eq. (2)] we must solve this equation. It can be shown (proof below) that the solution of Eq. (6) can be written as 
\begin{eqnarray}
\v E_s=-\frac{1}{4\pi}\int\int d^3x'dt'G_R\bigg(\nabla'\rho+\frac{1}{c^2}\frac{\partial \v j}{\partial t'}\bigg)-\v E_i,
\end{eqnarray}
where $G_R= \delta(t'-t+R/c)/R$ is the retarded Green function
satisfying $\Box^2 G_R(\v x,t;\v x',t') = -4\pi\delta(\v x-\v x')\delta(t-t')$ with $\Box^2\equiv\nabla^2-(1/c^2)\partial^2/\partial t^2$ being the D'Alambertian operator. As may be seen, Eq. (7) contains the term $-\v E_i$ and therefore the irrotational component $\v E_i$ appearing in the electric field $\v E=\v E_i+\v E_s$ is exactly canceled by the term $-\v E_i$ appearing in the solenoidal component $\v E_s$ given by Eq. (7). The fact that $\v E_i$ is an instantaneous component is physically irrelevant because it is always eliminated. Any possible interaction of $\v E_i$ with a charge $e$, for example, that given by the force $e\v E_i$ is automatically eliminated in the Lorentz force $\v F=e\v E_i+e\v E_s=e\v E_i + e\v E-e\v E_i=e\v E$. 
In other words: the field $\v E_i$ is a spurious field and therefore physically undetectable.

After substituting Eq. (7) into $\v E=\v E_i+\v E_s$, we obtain the usual retarded solution of Maxwell's equations for the electric field:
\begin{eqnarray}
\v E=-\frac{1}{4\pi}\int\int d^3x'dt'G_R\bigg(\nabla'\rho+\frac{1}{c^2}\frac{\partial \v j}{\partial t'}\bigg),
\end{eqnarray}
which propagates with the speed $c$. We can now answer the question: How does the field $\v E_s$ propagate? Answer: The field $\v E_s$ in Eq. (7) contains two parts, one of which propagates with the speed $c$ [the first term] and the other one with infinite speed [the second term] which is always canceled by the irrotational component $\v E_i$. This means that causality is never effectively lost in applying the Helmholtz theorem to the electric field of Maxwell's equations.

As above stated the instantaneous component $\v E_i$ is a spurious quantity which has mathematical but not physical existence. This result emphasizes the fact that the standard Helmholtz decomposition of the electric field involves terms with no physical significance. Yang [7]  has recently emphasized the difficulties arising from applying the Helmholtz theorem to time-dependent vector fields. He wrote [7]: ``There are two physics-related problems with this [Helmholtz] decomposition that are relevant: It introduces a spurious nonlocal property and spurious propagation behavior into the gradient and curl components. The result of the Helmholtz theorem are not physically consistent with the original vector function because of the spurious properties of its components."
He observes that in the Helmholtz decomposition of the current electric $\v j=\v j_i+\v j_s$, the components $\v j_i$ and $\v j_s$ do not in general vanish outside the source region and so they cannot be physically measured. He also notes that in the Helmholtz decomposition of the Lorenz-gauge vector potential  
$\v A^L=\v A^L_i+\v A^L_s$, the components $\v A^L_i$ and $\v A^L_s$ propagate ahead of its progenitor $\v A^L.$

It is easy to show that Eq. (7) satisfies Eq. (6). We simply take the D'Alambertian to Eq. (7) to obtain 
\begin{eqnarray}
\Box^2\v E_s= 4\pi\nabla\rho+\frac{4\pi}{c^2}\frac{\partial \v j}{\partial t}-\nabla^2\v E_i+ \frac{1}{c^2}\frac{\partial^2 \v E_i}{\partial t^2}.
\end{eqnarray}
If we use Eq. (1) then Eq. (9) reduces to 
\begin{eqnarray}
\Box^2\v E_s=\frac{4\pi}{c^2}\frac{\partial \v j}{\partial t}+ \frac{1}{c^2}\frac{\partial^2 \v E_i}{\partial t^2},
\end{eqnarray}
which is the same as Eq. (6). Furthermore, if we use Eq. (5) and $\v j=\v j_i-\v j_s$ then Eq. (10) becomes
\begin{eqnarray}
\Box^2\v E_s=\frac{4\pi}{c^2}\frac{\partial \v j_s}{\partial t},
\end{eqnarray}
which is the same as Eq. (2). Alternatively, we can integrate Eq. (10) to obtain the solution (7). This procedure is somewhat laborious. In fact, the solution of (10) can be written as
\begin{eqnarray}
\v E_s=-\frac{1}{4\pi c^2}\int\int d^3x'dt'G_R\frac{\partial \v j}{\partial t'}
-\frac{1}{4\pi c^2}\int\int d^3x'dt'G_R\frac{\partial^2 \v E_i}{\partial t'^2}.
\end{eqnarray}
The second term on the right-hand side of Eq. (12) can be transformed using the tensor identity:
\begin{eqnarray}
-\frac{1}{c^2}G_R\frac{\partial^2E^j}{\partial t'^2}=-
G_R\partial'_k\partial'^kE^j+E^j\bigg(\partial'_k\partial'^k-\frac{1}{c^2}\frac{\partial^2}{\partial t'^2}\bigg)G_R \nonumber\\+\partial'_k(G_R\partial'^kE^j-E^j\partial'^k G_R)\qquad \qquad\quad\;\;\,
\nonumber\\-\frac{1}{c^2}\frac{\partial}{\partial t'}\bigg(G_R \frac{\partial E^j}{\partial t'} -E^j\frac{\partial G_R}{\partial t'}\bigg),\quad\quad\qquad\,\,
\end{eqnarray}
where $E^j=(\v E_i)^j$ and $\partial'^j=(\nabla')^j$. Latin indices $k$ and $j$ run from 1 to 3 and the summation convention on repeated indices is adopted. Integrating Eq. (13) over all space and all time (from $t=-\infty$ to $t=\infty$) and using $\partial'_k\partial'^k E^j=4\pi\partial^j\rho$ and $(\partial'_k\partial'^k-(1/c^2)\partial^2/\partial t'^2)G_R=-4\pi\delta(x^j-x'^j)\delta(t-t')$, we obtain 
\begin{eqnarray}
-\frac{1}{ c^2}\int\int d^3x'dt'G_R\frac{\partial^2 E^j}{\partial t'^2}\qquad\qquad\qquad\qquad\qquad\qquad\qquad\qquad\nonumber\\
=-\int d^3x'dt'G_R \partial'^j\rho -4\pi E^j\qquad\qquad\qquad\qquad\;\nonumber\\
+\int dt'\bigg\{\int d^3x'\partial'_k(G_R\partial'^k E^j-E^j\partial'^k G_R)\bigg\}\quad\;\,\,
\nonumber\\-\frac{1}{ c^2}\int d^3x'\bigg\{\int dt'\frac{\partial}{\partial t'}\bigg(G_R \frac{\partial E^j}{\partial t'} -E^j\frac{\partial G_R}{\partial t'}\bigg)\bigg\}\;.
\end{eqnarray}
The volume integral within the third term on the right-hand side of Eq. (14) can be transformed into a surface integral which vanishes on account of the behavior of both $G_R$ and $E^i$ at spatial infinity. Similarly, the exact time integration within the fourth term on the right-hand side vanishes on account of the behavior of $G_R$ and $E^j$ at temporal infinity. Thus Eq. (14) reduces to an expression that multiplied by $[1/(4\pi)]$ takes the form 
\begin{eqnarray}
-\frac{1}{4\pi c^2}\int\int d^3x'dt'G_R\frac{\partial^2 E^j }{\partial t'^2}=
-\frac{1}{4\pi}\int d^3x'dt'G_R \partial'^j\rho -E^j,
 \end{eqnarray}
or equivalently,
\begin{eqnarray}
-\frac{1}{4\pi c^2}\int\int d^3x'dt'G_R\frac{\partial^2\v E_i }{\partial t'^2}= -\frac{1}{4\pi}\int d^3x'dt'G_R\nabla'\rho -\v E_i.
 \end{eqnarray}
From Eqs. (12) and (16) we obtain the solution of (7).
\bigskip
\bigskip

\noindent {\normalsize {\bf 3. TWO MECHANISMS OF TRANSMISSION?}}

\bigskip
\noindent
The conclusion of Chubykalo et al [2] that there are two mechanisms of transmission of energy and momentum in classical electrodynamics, one instantaneous (via $\v E_i$) and the other one retarded (via $\v E_s$ and $\v B$), has no basis because the mechanism of instantaneous transmission of $\v E_i$ is canceled by a part of $\v E_s$. The remaining part of $\v E_s$ is the well-known retarded contribution. The final result is that there is only one mechanism of transmission in electrodynamics of localized sources in vacuum: the mechanism associated with the retarded fields propagating with the speed $c$. Whatever propagation or nonpropagation characteristics are exhibited by the irrotational and solenoidal components of the electric field, this field always display the experimentally verified properties of causality and propagation at speed of light $c$.  

The supposed necessity of considering both instantaneous and retarded interactions in classical electrodynamics was previously suggested by Chubykalo and Vlaev [4] on the basis of a supposed failure of the Lienard-Wiechert fields to satisfy the Maxwell equations. The demonstration of the supposed failure was proved to be false by Jackson [3] in a comment on the paper of Chubykalo and Vlaev [4] (an unconvincing reply [8] of these authors has not been published in the pertinent Journal). Jackson wrote [3]: ``It is of course known that in certain gauges the potentials can contain both retarded and instantaneous contributions. But there is {\it no necessity} for such a mixture. And the fields are always retarded..." He also pointed out [3]: ``Classical electromagnetic theory is complete as usually expressed. One may choose to work in the Lorenz gauge in which all quantities are retarded." The comment of Jackson [3] was completely ignored in the paper of Chubykalo et al [2]. 

Paraphrasing Jackson, we can say that it is known that the standard Helmholtz decomposition of the electric field involves instantaneous and retarded components but there is no necessity of considering such a decomposition. After all, the full electric field is always retarded and one may choose to use a retarded Helmholtz's decomposition in which all quantities are retarded.
We will discuss this in section 5.

\bigskip
\bigskip

\noindent {\normalsize {\bf 4. THE WRONG EXAMPLE}}

\bigskip
\noindent
Chubykalo et al [2] claim: ``Let us consider the case when exclusively $\v E_i$ can be responsible for a signal transfer from a point charge q to the other point charge Q..." Their argument is as follows. They consider an arbitrarily moving charge for which $\rho(\v r,t)=q\delta(\v r-\v r_q(t))$ and $\v j(\v r,t)=\v v_q\delta(\v r-\v r_q(t))$, where $\v r_q(t)$ and $\v v_q(t)$ are the position and velocity of the particle. From Eqs. (30) and (42) of their paper [2] they obtain
\begin{eqnarray}
\v E_i=q \frac{\v r-\v r_q(t)}{|\v r-\v r_q(t)|^3},
\end{eqnarray} 
[Eq. (49) of Ref. 2], and state that this field [2]: ``...is a Coulomb type field: it is conservative ..." They then consider the particular case of a point charge oscillating along the X-axis: $\v r_q=(A_0\sin \omega t)\v i$, and find
\begin{eqnarray}
\nabla\times\v E_i=0,
\end{eqnarray}
[Eq. (55) of Ref. 2]. After some considerations, they conclude [2]: ``On account of the symmetry of the problem and because of  $\v E=\v E_i+\v E_s,\, \v E_s$ must be equal to $zero$ along of the X-axis. It can mean solely the following: The irrotational component of the electric field has a physical meaning and in some case is charged with the instantaneous energy and momentum transmission."  In a few words, they pretend to have found an example in which the full field $\v E$ equals its irrotational component $\v E_i$ along the X-axis:
\begin{eqnarray}
\v E=\v E_i. 
\end{eqnarray}
There are several ways to prove the inconsistence of Eq. (19) in Maxwell's theory. The simplest way is the obvious one: Any time-dependent electric field $\v E$ of Maxwell's theory must satisfy Faraday's law: 
\begin{eqnarray}
\nabla\times\v E=-\frac{1}{c}\frac{\partial\v B}{\partial t}. 
\end{eqnarray}
But the field $\v E=\v E_i$ satisfies Eq. (18) and then Faraday's law is not obeyed! Moreover, if for a moment we would consider the possibility that $\v E_s=0$, as Chubykalo et al [2] claim to have proved in their particular example, then the solution (7) leads to 
\begin{eqnarray}
\v E_i=-\frac{1}{4\pi}\int\int d^3x'dt'G_R\bigg(\nabla'\rho+\frac{1}{c^2}\frac{\partial \v j}{\partial t'}\bigg).
\end{eqnarray}
The inconsistence of Eq. (21) becomes evident for the case of an arbitrarily moving charge. In this case the left-hand side of Eq. (21) gives the instantaneous field in Eq. (17) while the right-hand side gives  
the Lienard-Wiechert electric field, {\it i.e.,}
\begin{eqnarray}
q \frac{\v n}{R^2}= q
\Bigg[\frac{\v n-\v v_q/c}{\gamma^2(1-\v n\cdot\v v_q/c)^3R^2}\Bigg]_{\rm ret}\qquad\;\;\nonumber\\ \;+\frac{q}{c}\Bigg[\frac{\v n\times \{(\v n-\v v_q/c)\times \v a_q/c}{(1-\v n\cdot\v v_q/c)^3R}\Bigg]_{\rm ret},
\end{eqnarray} 
where the subscript ret means that the quantity enclosed in the square brackets is to be evaluated at the retarded time
$t'=t-R(t')/c;\, \v n=\v R/R=\v r-\v r_q(t)/|\v r-\v r_q(t)|$ and $\gamma= (1-v_q^2/c^2).^{-1/2}$ A simple reflection shows that Eq. (22) is manifestly inconsistent. The right-hand side cannot be equal to the left-hand side. The choice $c\to \infty$ is physically forbidden in the right-hand side of Eq. (22).

As a consolation for the authors of Ref. 2, their claim that in some cases the full electric field can equal to its irrotational part is shown to be correct in the context of a Galilean-invariant electromagnetic theory whose field equations are   [5,6]:  
\begin{eqnarray}
\nabla\cdot\widetilde{\v E}= 4\pi\rho,\\ 
\nabla\cdot\widetilde{\v B}=0,\;\;\, \\
\nabla\times\widetilde{\v E}=0,\;\;\,\\ 
\nabla\times\widetilde{\v B}-\frac{1}{c}\frac{\partial\widetilde{\v E}}{\partial t}=\frac{4\pi}{c}\v j,
\end{eqnarray}
where $\widetilde{\v E}$ and $\widetilde{\v B}$ are instantaneous electric and magnetic fields. From Eqs. (23) and (25) we obtain 
$\nabla^2\widetilde{\v E}=4\pi\nabla\rho$. Using the Helmholtz's theorem we have $\widetilde{\v E}=\widetilde{\v E}_i$. It follows that $\nabla^2\widetilde{\v E}_i=4\pi\nabla\rho$. The solution of this equation for $\rho(\v r,t)=q\delta(\v r-\v r_q(t))$ gives naturally Eq. (17):
\begin{eqnarray}
\widetilde {\v E}_i=q \frac{\v r-\v r_q(t)}{|\v r-\v r_q(t)|^3}.
\end{eqnarray} 

\bigskip
\bigskip
\noindent {\normalsize  {\bf 5. THE SUPPOSED INFERENCE}}
\bigskip

\noindent
Chubykalo et al [2]: ``... although the electric field... can be {\it retarded}, it is decomposed into just two parts, one of which is {\it pure irrotational} and the other is {\it pure solenoidal}:
\begin{eqnarray}
\v E=\v E_i+\v E_s, \quad \v E_i=-\nabla\varphi, \quad \v E_s=-\frac{1}{c}\frac{\partial\v A}{\partial t},
\end{eqnarray} 
(in the Coulomb gauge $\nabla\cdot \v A=0)$ [Eq. (6) of Ref. 2]. This alone shows that {\bf the inference of J. A. Heras} ... that a retarded field cannot be decomposed into {\it only} two parts (irrotational and solenoidal) can be insufficiently rigorous." The boldface emphasis is mine. First of all, I have never made this incorrect 
inference as Chubykalo et al [2] claim. Some years ago McQuistan [9] and more recently the present author [10-12], have formulated the retarded Helmholtz theorem which states that a retarded field vanishing at infinity can be decomposed into irrotational, solenoidal and temporal components: 
\begin{eqnarray}
\v E=\v E^{\cal R}_i+\v E^{\cal R}_s +\v E^{\cal R}_{\cal T},
\end{eqnarray} 
where 
\begin{eqnarray}
\v E^{\cal R}_i=-\nabla\int d^3x'\frac{[\nabla'\cdot\v E]}{4\pi R},\;\;\,\\
\v E^{\cal R}_s=\nabla\times\int d^3x'\frac{[\nabla'\times\v E]}{4\pi R},\\
\v E^{\cal R}_{\cal T}= \frac{1}{c^2}\frac{\partial}{\partial t}\int d^3x'\frac{[\partial \v E/\partial t]}{4\pi R}.
\end{eqnarray}
The square brackets $[\;]$ mean that the enclosed quantity is to be evaluated at the retarded time $t'=t-R/c$,
and the superscript ${\cal R}$ emphasizes the retarded character of the quantities [13]. An alternative version of this retarded Helmholtz theorem has also recently formulated [14]. The result that the field $\v E$ can be decomposed in terms of the components $\v E^{\cal R}_i,\,\v E^{\cal R}_s$ and $ \v E^{\cal R}_{\cal T}$ does not exclude the possibility that $\v E$ can also be decomposed in terms of other different irrotational and solenoidal components $\v E_i$ and $\v E_s$. In Ref. 12, I show that if $\v E$ is the retarded electric field of Maxwell's equations then 
\begin{eqnarray}
\v E^{\cal R}_i=-\nabla\int d^3x'\frac{[\rho]}{4\pi R},\,\\
\v E^{\cal R}_s+\v E^R_{\cal T}=- \frac 1c\frac{\partial}{\partial t} \int d^3x' \frac{[\v j]}{Rc},
\end{eqnarray}
and I show also that these expressions can be written in terms of the Coulomb-gauge potentials $\Phi_C$ and $\v A_C$ as follows:
\begin{eqnarray}
\v E^{\cal R}_i=-\nabla\Phi_C+\frac{1}{c^2}\frac{\partial^2}{\partial t^2}\int d^3x'\frac{[\nabla'\Phi_C]}{4\pi R},\;\;\,\\
\v E^{\cal R}_s+\v E^R_{\cal T}=-\frac{1}{c}\frac{\partial\v A_C}{\partial t}-\frac{1}{c^2}\frac{\partial^2}{\partial t^2}\int d^3x'\frac{[\nabla'\Phi_C]}{4\pi R}.
\end{eqnarray}
Therefore,
\begin{eqnarray} 
\v E=\v E^{\cal R}_i+\v E^{\cal R}_s +\v E^{\cal R}_{\cal T}=-\nabla\Phi_C-\frac{1}{c}\frac{\partial\v A_C}{\partial t}.
\end{eqnarray}
This result means  that the retarded electric field can $rigorously$ be decomposed either in terms of the irrotational, solenoidal and temporal components: $\v E^{\cal R}_i,\, \v E^{\cal R}_s$ and $\v E^{\cal R}_{\cal T}$, or equivalently in terms of the instantaneous irrotational component: $-\nabla\Phi_C$ and of the solenoidal component: $-(1/c)\partial\v A_C/\partial t.$ 
I have never inferred that: ``... a retarded field cannot be decomposed into {\it only} two parts (irrotational and solenoidal)..." as Chubykalo et al [2] attribute to me. Furthermore, I have proved in Ref. 12 exactly the opposite: By applying the retarded Helmholtz theorem, I could decompose the retarded electric field into two parts (irrotational and solenoidal) as may be seen in Eq. (37). 

The direct application of the retarded Helmholtz theorem to the electric field of Maxwell's equations leads to the well-known retarded expression of this field [see Eqs. (33) and (34)] which of course do not include instantaneous contributions. On the other hand, the direct application of the standard Helmholtz theorem to the electric field of Maxwell's equations leads to the expression of this field in terms of the Coulomb-gauge potentials [see Eq. (6) of Ref. 2 or Eq. (28) in the present paper] with the disadvantage that in such an application  a spurious instantaneous electric field is introduced. Accordingly, if we do not want to generally introduce instantaneous fields using the standard Helmholtz theorem then we may use the retarded form of this theorem [10-12].

Chubykalo et al [2] attempt: `` ... to substantiate the applying of the Helmholtz theorem to vector fields in classical electrodynamics." Unfortunately the physical interpretations given by these authors for the irrotational and solenoidal components of the electric field, obtained from applying the standard Helmholtz theorem, are misleading and add nothing but confusion to the topic of instantaneous and retarded fields. The present author has formally demonstrated  [6] that the instantaneous fields can be introduced as {\it unphysical} objects into classical electrodynamics which can be used to express the retarded fields. 

\bigskip

\noindent {\bf Acknowledgements.} The present author is grateful to Professor R. F. O'Connell for the kind hospitality extended to him in    
the Department of Physics and Astronomy of the Louisiana State University.


\begin{thebibliography}{99}

\bibitem{1}  
J. D. Jackson, ``From Lorenz to Coulomb and other explicit gauge transformations," {\it Am. J. Phys.} {\bf 70}, 917 (2002).

\bibitem{2}
A. Chubykalo, A. Espinoza, R. Alvarado Flores, and A. Gutierrez Rodriguez, ``Helmholtz theorem and the v-gauge in the problems of superluminal and instantaneous signals in classical electrodynamics," {\it Found. Phys. Lett.} {\bf 19}, 37 (2006).

\bibitem{3}  
J. D. Jackson, ``Criticism of `Necessity of simultaneous co-existence of instantaneous and retarded interactions in classical electrodynamics, by Chubykalo and Vlaev,'" {\it Int. J. Mod. Phys. A} {\bf 17}, 3975 (2002).

\bibitem{4} 
A. Chubykalo and S. J. Vlaev, ``Necessity of simultaneous co-existence of instantaneous and retarded interactions in classical electrodynamics, by Chubykalo and Vlaev," {\it Int. J. Mod. Phys. A} {\bf 14}, 3789 (1999).

\bibitem{5} 
M. Jammer and J. Stachel,``If Maxwell had worked between Ampere and Faraday: An historical fable with a pedagogical moral," {\it Am. J. Phys.} {\bf 48}, 5 (1980).

\bibitem{6}  
J. A. Heras, ``Instantaneous fields in classical electrodynamics," {\it Europhys. Letts.} {\bf 69}, 1 (2005).  

\bibitem{7}
Kuo-Ho Yang, ``The physics of gauge transformations," {\it Am. J. Phys.} 
{\bf 73}, 742 (2005).

\bibitem{8}
A. Chubykalo and S. J. Vlaev, 
``Reply to 'Criticism of `Necessity of simultaneous co-Existence of instantaneous and retarded interactions in classical electrodynamics' by J.D.Jackson," e-print: physics/0205041 

\bibitem{9}
R. B. McQuistan, {\it Scalar and Vector fields: A Physical Interpretation} (Wiley, New York, 1965), Sec. 12-3. 

\bibitem{10}  
 J. A. Heras, ``Jefimenko's formulas with magnetic monopoles and the Lienard-Wiechert fields of a dual-charged particle," {\it Am. J. Phys.} {\bf 62}, 525 (1994).

\bibitem{11}  
 J. A. Heras, ``Time-dependent generalizations of the 
Biot-Savart and Coulomb laws: A formal derivation," {\it Am. J. Phys.} {\bf 63}, 928 (1995). 

\bibitem{12}  
 J. A. Heras, ``Comment on `Causality, the Coulomb field, and Newton's law of gravitation' by F. Rohrlich 
[Am. J. Phys. {\bf 70}, 411-414 (2002)],'" {\it Am. J. Phys.} {\bf 71}, 729 (2003).

\bibitem{13} 
I pointed out in Refs. 10 and 11 that the retarded Helmholtz theorem [Eq. (29) in the present comment] is not very useful in practice because we cannot generally specify $\partial \v E/\partial t$. Despite of this objection, I show in Ref. 10 how the retarded Helmholtz theorem can successfully be applied to find the retarded solutions of Maxwell's equations for the general case that include magnetic monopoles. Also, I show in Ref. 11 how an equivalent form of theorem [Eq. (17) of Ref. 11] leads to the retarded solutions of Maxwell's equations expressed in the form of time-dependent generalizations of the Coulomb and Biot-Savart laws. 

\bibitem{14}
A. M. Davis, ``A generalized Helmholtz theorem for time-varying vector fields," {\it Am. J. Phys.} {\bf 74}, 72 (2006);
J. A. Heras,``Coment on `A generalized Helmholtz theorem for time-varying vector fields by A. M. Davis, [Am. J. Phys. {\bf 74}, 72 (2006)],'"                  {\it Am. J. Phys.} {\bf 74}, 743 (2006).

\end{thebibliography}
\end{document}